\documentclass[a4paper,11pt]{article}
\pdfoutput=1 

\usepackage{jheppub} 

\usepackage[T1]{fontenc} 
\usepackage{subcaption}

\title{How to Succeed at Witten Diagram Recursions without Really Trying}

\author[]{Xinan Zhou}
\affiliation[]{Princeton Center for Theoretical Science, Princeton University, Princeton, NJ 08544, USA }

\emailAdd{xinanz@princeton.edu}

\abstract{Witten diagrams are basic objects for studying dynamics in AdS space, and also play key roles in the analytic functional bootstrap. However, these diagrams are notoriously hard to evaluate, making it extremely difficult to search for recursion relations among them.  In this note, we present simple methods to obtain recursion relations for exchange Witten diagrams from conformal block recursion relations. We discover a variety of new relations, including the dimensional reduction formulae for exchange Witten diagrams. In particular, we find a five-term recursion relation relating exchange Witten diagrams in $d$ and $d-2$ dimensions. This gives the holographic analogue of a similar formula for conformal blocks due to Parisi-Sourlas supersymmetry. We also extend the analysis to two-point functions in CFTs with conformal boundaries, and obtain similar results.  }

\begin{document}
\maketitle
\flushbottom
\section{Introduction}
Concrete results on conformal blocks paved the way for the modern return of the conformal bootstrap program.\footnote{See, {\it e.g.}, \cite{Rychkov:2016iqz,Simmons-Duffin:2016gjk} for pedagogical lecture notes.} First appeared in the 1970s \cite{Polyakov:1974gs,Ferrara:1973vz,Ferrara:1974nf,Ferrara:1974ny}, these objects did not receive much attention until the breakthrough results by Dolan and Osborn \cite{Dolan:2000ut,Dolan:2003hv,Dolan:2011dv}, which were crucial for the development of the numerical bootstrap techniques \cite{Rattazzi:2008pe}. Since then conformal blocks have been intensively studied, and a great deal of beautiful properties have been discovered (see \cite{Poland:2018epd} for a review and references therein). In these results of conformal blocks, various recursion relations often play an important role. For example, conformal blocks in different spacetime dimensions can be recursively related \cite{Hogervorst:2016hal,Kaviraj:2019tbg}. Moreover, recursion relations also provide an efficient way to obtain spinning conformal blocks from scalar ones \cite{Costa:2011dw,Iliesiu:2015qra,Echeverri:2015rwa,Karateev:2017jgd}.

Operator exchange in conformal blocks, via holography, can naturally be  associated to the tree-level exchange process of a single-particle state in AdS space. The latter is characterized by exchange Witten diagrams, and are the building blocks for holographic computation of boundary correlators. Though similar intuitively, the two objects are  different in details.\footnote{The exact holographic dual of conformal blocks is the so-called ``geodesic'' Witten diagrams \cite{Hijano:2015zsa}. However, these objects do not naturally appear in the holographic calculation except in the semiclassical limit.} Under conformal block decomposition, the exchange Witten diagram contains a ``single-trace'' conformal block  with the same conformal dimension and spin. Additionally, the single-trace conformal block is dressed with  infinitely many ``double-trace'' conformal blocks which are bilinears of external operators. These double-trace conformal blocks are in fact far from random. For example, their OPE coefficients are fine-tuned to ensure that the Witten diagram is single-valued in Euclidean signature. While there has already  been a wealth of results on recursion relations for conformal blocks, little is known about the recursion relations for Witten diagrams\footnote{See however  \cite{Raju:2011mp,Raju:2010by,Raju:2012zr} for recursion relations for gravity and Yang-Mills in AdS, and \cite{Goncalves:2014rfa} for relating Witten diagrams to lower-point ones using factorization.}, partly due to the notorious difficulty in computing them \cite{Freedman:1998tz,Liu:1998ty,DHoker:1998ecp,DHoker:1999pj,DHoker:1999mqo,Hoffmann:2000tr,Hoffmann:2000mx,Arutyunov:2002fh,Bekaert:2014cea,Aharony:2016dwx,Yuan:2017vgp,Yuan:2018qva,Carmi:2019ocp,Meltzer:2019nbs} \footnote{To date, there are no closed form formulae for exchange Witten diagrams in terms elementary functions, except for special cases.}. Finding new relations among Witten diagrams, however, will no doubt be extremely useful. Such relations will facilitate various holographic calculations, and also shed light on new structures in scattering amplitudes in AdS. Moreover,  the actions of analytic functionals are succinctly encoded in exchange Witten diagrams \cite{Mazac:2018ycv,Mazac:2018biw,Mazac:2019shk}. We can therefore expect to extract a lot of lessons from the Witten diagram relations about the structures of the analytic functionals.

In this note, we accomplish the task of finding new Witten diagram relations by transferring our knowledge of conformal blocks. We give straightforward recipes to obtain a large class of recursion relations for Witten diagrams from those of conformal blocks. In the simplest scenario, the recipe amounts to just replacing the conformal blocks with the corresponding exchange Witten diagrams. This most basic version applies to recursion relations in which conformal blocks appear linearly with constant coefficients  independent of the cross ratios. Notable examples in this family include the dimensional reduction formulae
\begin{equation}\label{dtodm1}
g^{(d)}_{\Delta,\ell}=\sum_{n=0}^\infty\sum_{j} A_{n,j}\, g^{(d-1)}_{\Delta+2n,j}\;,\quad j=\ell\;,\ell-2\;,\ldots\;, \ell\;\text{mod}\;2\;,
\end{equation}
relating conformal blocks $d$ and $d-1$ dimensions \cite{Hogervorst:2016hal}, and 
\begin{equation}\label{PSsymmetry}
g^{(d-2)}_{\Delta,\ell}=g^{(d)}_{\Delta,\ell}+c_{2,0}g^{(d)}_{\Delta+2,\ell}+c_{1,-1}g^{(d)}_{\Delta+1,\ell-1}+c_{0,-2}g^{(d)}_{\Delta,\ell-2}+c_{2,-2}g^{(d)}_{\Delta+2,\ell-2}\;,
\end{equation}
relating $d$ and $d-2$ dimensions \cite{Kaviraj:2019tbg}. The  expressions for  constants $A_{n,j}$, $c_{i,j}$ are given in (\ref{Anj}), (\ref{cij}), but are inconsequential for the discussion. We claim that these identities correspondingly yield the following recursion relations for exchange Witten diagrams  $W^{(d)}_{\Delta,\ell}$ defined in $AdS_{d+1}$
\begin{equation}
W^{(d)}_{\Delta,\ell}=\sum_{n=0}^\infty\sum_{j} A_{n,j}\, W^{(d-1)}_{\Delta+2n,j}\;,\quad j=\ell\;,\ell-2\;,\ldots\;, \ell\;\text{mod}\;2\;,
\end{equation}
\begin{equation}
W^{(d-2)}_{\Delta,\ell}=W^{(d)}_{\Delta,\ell}+c_{2,0}W^{(d)}_{\Delta+2,\ell}+c_{1,-1}W^{(d)}_{\Delta+1,\ell-1}+c_{0,-2}W^{(d)}_{\Delta,\ell-2}+c_{2,-2}W^{(d)}_{\Delta+2,\ell-2}\;.
\end{equation}
These relations constitute the dimensional reduction formulae which relate exchange Witten diagrams from different bulk dimensions. Note that, importantly, exchange Witten diagrams with spin $\ell\geq 1$ are  well-defined only up to adding contact Witten diagrams with at most $2(\ell-1)$ derivatives. The ambiguity reflects different choices of the cubic vertices, which do not affect the single-trace operator exchange \cite{Costa:2014kfa}. Therefore, the above identities are valid when a proper choice of contact terms has been made (and there are generally infinitely many choices as we will see). These formulae may strike as odd, as they require miraculous cancellations of the double-trace operators. However, we can give simple arguments which explain why it can always happen. To see this, let us go to the Mellin space \cite{Mack:2009mi,Penedones:2010ue}.\footnote{Another way to prove these relations is to use the Lorentzian inversion formula \cite{Caron-Huot:2017vep,Simmons-Duffin:2017nub}. See Section \ref{Sec:2} for details.}
The Mellin amplitudes of the exchange Witten diagrams have the following simple structure
\begin{equation}\label{MellinW}
\mathcal{M}^{(d)}_{\Delta,\ell}(s,t)=\sum_{m=0}^\infty \frac{Q_{\ell,m}^{(d)}(t)}{s-(\Delta-\ell)-2m}+P^{(d)}_{\ell-1}(s,t)
\end{equation}
where $Q_{\ell,m}^{(d)}(t)$ and $P^{(d)}_{\ell-1}(s,t)$ are polynomials in $t$ and $s$, $t$  of degrees $\ell$ and $\ell-1$ respectively.  A well known property of the Mellin representation is that the Mellin amplitudes of conformal blocks and exchange Witten diagrams have the {\it same} poles and residues. Therefore, the transition from conformal blocks to Witten diagrams for the recursion relations boils down to merely tuning the regular pieces $P^{(d)}_{\ell-1}(s,t)$ in each diagram. This turns out to be always possible, thanks to the contact term ambiguity.  A particularly simple choice for $W^{(d)}_{\Delta,\ell}$ is the ``Polyakov-Regge'' blocks introduced in \cite{Mazac:2019shk,Sleight:2019ive}, which correspond to setting all $P^{(d)}_{\ell-1}(s,t)$ to be zero. 

The above recipe also extends to recursion relations involving differential operators. For certain well-behaved operators, the recipe is only modified by adding higher-derivative contact Witten diagrams. To see this, we note that differential operators can be interpreted as difference operators acting on Mellin amplitudes. Let us consider for now difference operators which do not introduce new poles. 
Then by construction the polar part of the Mellin amplitudes on both sides are automatically matched, and we are left with only regular polynomial terms. Additional contact terms are required if the regular terms have degrees too high to be absorbed into the exchange Witten diagrams. A case in point is the Casimir equation for conformal blocks \cite{Dolan:2003hv}
\begin{equation}\label{Casimireqn}
\mathrm{Cas}_s[g^{(d)}_{\Delta,\ell}]-\mathcal{C}_{\Delta,\ell}\,g^{(d)}_{\Delta,\ell}=0
\end{equation}
where $\mathcal{C}_{\Delta,\ell}=\Delta(\Delta-d)+\ell(\ell+d-2)$ is the eigenvalue of the Casimir. The corresponding equation for Witten diagrams is the well known equation of motion identity
\begin{equation}\label{EOMid}
\mathrm{Cas}_s[W^{(d)}_{\Delta,\ell}]-\mathcal{C}_{\Delta,\ell}\,W^{(d)}_{\Delta,\ell}=W^{(d)}_{con}\;.
\end{equation}
Note that $W^{(d)}_{con}$ is a contact diagram with $2\ell$ derivatives. It cannot be absorbed into  the exchange Witten diagram $W^{(d)}_{\Delta,\ell}$, which is ambiguous up to contact diagrams with only no more than $2(\ell-1)$ derivatives. Further extensions including more general differential operators are also possible, as we will discuss in the main text of the paper. Such generalizations generically require adding extra diagrams which have singularities. 

The rest of the paper is organized as follows. In Section \ref{Sec:2}, we elaborate on the dimensional reduction for Witten diagrams, and comment on a number of interesting properties. In Section \ref{Sec:3} we discuss the modification of the recipe for more general recursion relations. They include in particular the action of the crossed-channel conformal Casimir, weight-shifting operators and superconformal Ward identities. We also discuss the generalization to CFTs with conformal boundaries in Section \ref{Sec:4}. We give the dimensional reduction formulae for BCFT conformal blocks in $d$ dimensions, as an infinite sum of conformal blocks in $d-1$ dimensions. We also find two-term relations analogous to (\ref{PSsymmetry}), which relate conformal blocks in $d$ and $d-2$ dimensions. Since the boundary channel conformal block coincides with the scalar bulk-to-bulk propagator in AdS, we can also reinterpret the results as the dimensional reduction of AdS propagators. The paper concludes in Section \ref{Sec:5} with a brief discussion of the results and outline for future directions. Various technical details are relegated to the two appendices, where we also work out a number of explicit examples.

\section{Dimensional reduction for Witten diagrams}\label{Sec:2}
In this section we continue to discuss the simplest scenario for obtaining Witten diagram recursion relations, where they descend directly from those of conformal blocks. This class of recursion relations take the form of linear combinations of conformal blocks with pure number coefficients.

We will focus on two representative recursion relations which relate conformal blocks from different spacetime dimensions. The first relation expresses $d$-dimensional conformal blocks in terms of infinitely many $(d-1)$-dimensional conformal blocks \cite{Hogervorst:2016hal}\footnote{In this paper, we use the same normalization for the conformal blocks as in \cite{Poland:2018epd}.}
\begin{equation}
g^{(d)}_{\Delta,\ell}=\sum_{n=0}^\infty\sum_{j} A_{n,j}\, g^{(d-1)}_{\Delta+2n,j}\;,\quad j=\ell\;,\ell-2\;,\ldots\;, \ell\;\text{mod}\;2\;,
\end{equation}
where 
\begin{equation}\label{Anj}
A_{n,j}=\frac{Z^j_\ell 16^{-n} \left(\frac{1}{2}\right)_n (\Delta -1)_{2 n} \left(\frac{j+\Delta }{2}\right)_n \left(\frac{\ell +\Delta }{2}\right)_n\left(\frac{-d-j+\Delta +3}{2}\right)_n \left(\frac{-d-\ell +\Delta +2}{2}\right)_n}{n! \left(\frac{2\Delta +2-d}{2}\right)_n \left(\frac{2n+2\Delta+1-d}{2}\right)_n\left(\frac{j+\Delta -1}{2}\right)_n \left(\frac{\ell +\Delta +1}{2}\right)_n \left(\frac{-d-j+\Delta +2}{2}\right)_n \left(\frac{-d-\ell +\Delta +3}{2}\right)_n}\;,
\end{equation}
\begin{equation}
Z^j_\ell=\frac{\ell ! (d+2 j-3) (d-3)_j \left(\frac{1}{2}\right)_{\frac{\ell -j}{2}} \left(\frac{d-2}{2}\right)_{\frac{j+\ell }{2}}}{2 j! \frac{\ell -j}{2}! (d-2)_{\ell } \left(\frac{d-3}{2}\right)_{\frac{1}{2} (j+\ell +2)}}\;.
\end{equation}
The second relation relates $d$ and $d-2$ dimensions \cite{Kaviraj:2019tbg}
\begin{equation}
g^{(d-2)}_{\Delta,\ell}=g^{(d)}_{\Delta,\ell}+c_{2,0}g^{(d)}_{\Delta+2,\ell}+c_{1,-1}g^{(d)}_{\Delta+1,\ell-1}+c_{0,-2}g^{(d)}_{\Delta,\ell-2}+c_{2,-2}g^{(d)}_{\Delta+2,\ell-2}\;,
\end{equation}
where the coefficients are
\begin{equation}
\begin{split}\label{cij}
{}&c_{2,0}=-\tfrac{(\Delta -1) \Delta  (\Delta -\Delta_{12}+\ell ) (\Delta +\Delta_{12}+\ell ) (\Delta -\Delta_{34}+\ell ) (\Delta +\Delta_{34}+\ell )}{4 (d-2 \Delta -4) (d-2 \Delta -2) (\Delta +\ell -1) (\Delta +\ell )^2 (\Delta +\ell +1)}\;,\\
{}&c_{1,-1}=-\tfrac{(\Delta -1) \Delta_{12} \Delta_{34} \ell }{(\Delta +\ell -2) (\Delta +\ell ) (d-\Delta +\ell -4) (d-\Delta +\ell -2)}\;,\\
{}&c_{0,-2}=-\tfrac{(\ell -1) \ell }{(d+2 \ell -6) (d+2 \ell -4)}\;,\\
{}&c_{2,-2}=\tfrac{(\Delta -1) \Delta  (\ell -1) \ell  (d-\Delta -\Delta_{12}+\ell -4) (d-\Delta +\Delta_{12}+\ell -4) (d-\Delta -\Delta_{34}+\ell -4) (d-\Delta +\Delta_{34}+\ell -4)}{4 (d-2 \Delta -4) (d-2 \Delta -2) (d+2 \ell -6) (d+2 \ell -4) (d-\Delta +\ell -5) (d-\Delta +\ell -4)^2 (d-\Delta +\ell -3)}\;,
\end{split}
\end{equation}
with $\Delta_{ij}=\Delta_i-\Delta_j$. That the relation contains only finitely many $d$-dimensional conformal blocks is a consequence of the underlying Parisi-Sourlas supersymmetry \cite{Parisi:1979ka}.

As we have argued in the introduction, we can obtain corresponding dimensional reduction formula by simply replacing $g^{(d)}_{\Delta,\ell}$ with the exchange Witten diagram $W^{(d)}_{\Delta,\ell}$
\begin{equation}\label{Wreddtodm1}
W^{(d)}_{\Delta,\ell}=\sum_{n=0}^\infty\sum_{j} A_{n,j}\, W^{(d-1)}_{\Delta+2n,j}\;,\quad j=\ell\;,\ell-2\;,\ldots\;, \ell\;\text{mod}\;2\;,
\end{equation}
\begin{equation}\label{Wreddtodm2}
W^{(d-2)}_{\Delta,\ell}=W^{(d)}_{\Delta,\ell}+c_{2,0}W^{(d)}_{\Delta+2,\ell}+c_{1,-1}W^{(d)}_{\Delta+1,\ell-1}+c_{0,-2}W^{(d)}_{\Delta,\ell-2}+c_{2,-2}W^{(d)}_{\Delta+2,\ell-2}\;.
\end{equation}
Note that the Witten diagrams $W^{(d)}_{\Delta,\ell}$ are normalized such that the single-trace conformal block appears with coefficient one. For the relations (\ref{Wreddtodm1}) and (\ref{Wreddtodm2}) to be valid, the contact part in the exchange Witten diagrams must be constrained. To precisely state these constraints, let us recall the definition for the Mellin amplitude \cite{Mack:2009mi,Penedones:2010ue}
\begin{equation}\label{defMellin}
G(x_i)=\frac{\left(\frac{x_{14}^2}{x_{24}^2}\right)^{\frac{\Delta_2-\Delta_1}{2}}\left(\frac{x_{14}^2}{x_{13}^2}\right)^{\frac{\Delta_3-\Delta_4}{2}}}{(x_{12}^2)^{\frac{\Delta_1+\Delta_2}{2}}(x_{34}^2)^{\frac{\Delta_3+\Delta_4}{2}}}\int \frac{dsdt}{(4\pi i)^2}U^{\frac{s}{2}} V^{\frac{t}{2}-\frac{\Delta_2+\Delta_3}{2}}\mathcal{M}(s,t)\Gamma_{\Delta_1\Delta_2\Delta_3\Delta_4}(s,t)
\end{equation}
where
\begin{equation}
U=\frac{x_{12}^2x_{34}^2}{x_{13}^2x_{24}^2}\;,\quad V=\frac{x_{14}^2x_{23}^2}{x_{13}^2x_{24}^2}\;,
\end{equation}
are the conformal cross ratios and
\begin{equation}
\Gamma_{\Delta_1\Delta_2\Delta_3\Delta_4}=\Gamma[\tfrac{\Delta_1+\Delta_2-s}{2}]\Gamma[\tfrac{\Delta_3+\Delta_4-s}{2}]\Gamma[\tfrac{\Delta_1+\Delta_4-t}{2}]\Gamma[\tfrac{\Delta_2+\Delta_3-t}{2}]\Gamma[\tfrac{\Delta_1+\Delta_3-u}{2}]\Gamma[\tfrac{\Delta_2+\Delta_4-u}{2}]\;,
\end{equation}
with $s+t+u=\Delta_1+\Delta_2+\Delta_3+\Delta_4$. Translating (\ref{Wreddtodm1}), (\ref{Wreddtodm2}) into Mellin space, the polar part on both sides clearly are already matched since their residues just reproduce the single-trace conformal blocks. The regular part is constrained by the conditions 
\begin{equation}\label{Pcond1}
P^{(d)}_{\Delta,\ell}(s,t)=\sum_{n=0}^\infty\sum_{j} A_{n,j}\, P^{(d-1)}_{\Delta+2n,j}(s,t)\;,
\end{equation}
\begin{equation}\label{Pcond2}
P^{(d-2)}_{\Delta,\ell}=P^{(d)}_{\Delta,\ell}+c_{2,0}P^{(d)}_{\Delta+2,\ell}+c_{1,-1}P^{(d)}_{\Delta+1,\ell-1}+c_{0,-2}P^{(d)}_{\Delta,\ell-2}+c_{2,-2}P^{(d)}_{\Delta+2,\ell-2}\;.
\end{equation}
On the other hand, the Mellin amplitude of a contact diagram with  $2L$ derivatives is a polynomial with degree $L$. Since $W^{(d)}_{\Delta,\ell}$ can absorb any contact Witten diagrams with no more than $2(\ell-1)$ derivatives by adjusting cubic couplings \cite{Costa:2014kfa}, $P^{(d)}_{\Delta,\ell}$ can be changed by adding any degree $\ell-1$ polynomials in $s$ and $t$. It is clear that the constraints (\ref{Pcond1}) and (\ref{Pcond2}) can always be solved. In fact, a particularly simple solution is to use the contact term ambiguity to remove all $P^{(d)}_{\Delta,\ell}$, and the constraints are trivially satisfied. These special Witten diagrams are uniquely defined, and were dubbed the Polyakov-Regge blocks in \cite{Mazac:2019shk,Sleight:2019ive}. They have enhanced Regge behavior in the t-channel
\begin{equation}
\mathcal{M}^{(d)}_{\Delta,\ell}(s,t)\propto \frac{1}{s}\;, \quad \quad s\to\infty\;,\; t\;\;{\rm fixed}\;. 
\end{equation}
While the existence of a finite-term reduction formula (\ref{PSsymmetry}) is already quite remarkable, thanks to the Parisi-Sourlas supersymmetry, it is even more nontrivial that a similar relation exists for AdS exchange diagrams. It shows that one can construct a different AdS effective Lagrangian (albeit non-unitary) in two dimensions higher, which produces the same tree-level scattering amplitude. It may also be of interest to spell out the details of how the Parisi-Sourlas supersymmetry is realized in the bulk.

Let us comment that the above relations can also be proven using the Lorentzian inversion formula \cite{Caron-Huot:2017vep,Simmons-Duffin:2017nub}. Since double-trace conformal blocks do not contribute to the double discontinuity, both sides of (\ref{Wreddtodm1}) and (\ref{Wreddtodm2}) have the same double discontinuities thanks to (\ref{dtodm1}) and (\ref{PSsymmetry}). By using the inversion formula, they lead to the same ``coefficient function'' $c(\Delta,J)$ for the conformal partial wave expansion in the crossed channel.\footnote{Note that we can use on both sides, {\it e.g.}, the inversion formula in $d$ dimensions. This does not give the the canonical decomposition on one side as the conformal partial waves are from a different dimension, but it does not cause any problem. } However, note that the Lorentzian inversion formula for a spin-$\ell$ exchange Witten diagram generally converges only to $J\geq \ell$. In order to reach zero spin, we can improve the Regge behavior of exchange Witten diagrams by adding contact terms and obtain the Polyakov-Regge blocks. Then the  inversion formula applies to all spins, and we have proven (\ref{Wreddtodm1}) and (\ref{Wreddtodm2}).

Let us also point out a few interesting features of the relation (\ref{Wreddtodm1}). When the dimension of the exchanged operator takes special values with respect to the external dimensions
\begin{equation}
\Delta_1+\Delta_2-\Delta=2n_0\in 2\mathbb{Z}_+\;,\quad\quad \text{or}\quad\quad \Delta_3+\Delta_4-\Delta=2n'_0\in 2\mathbb{Z}_+
\end{equation}
the infinite sum in (\ref{Wreddtodm1}) truncates to finitely many terms\footnote{The extra conformal blocks in (\ref{dtodm1}) however are not absent. Instead they now coincide with the double-trace conformal blocks in the exchange Witten diagrams.} 
\begin{equation}\label{Wreddtodm1trunc}
W^{(d)}_{\Delta,\ell}=\sum_{n=0}^{n_{\max}}\sum_{j} A_{n,j}\, W^{(d-1)}_{\Delta+2n,j}\;,\quad  j\geq \ell+2(n-n_{\max})
\end{equation}
with $n_{\max}=n_0-1$ or $n'_0-1$, or $\min\{n_0-1,n'_0-1\}$ if both conditions are satisfied. This is because the exchange Mellin amplitudes  (\ref{MellinW}) truncate to finitely many poles with $m_{\max}=n_{\max}$, and only finitely many diagrams are needed to match finitely many poles. The termination of Mellin poles was first explained in \cite{Rastelli:2017udc} as a consistency condition for large $N$ expansion. We will also give an explicit position space example of such finite-term recursion relation in Appendix \ref{App:B}. As a further special case of (\ref{Wreddtodm1trunc}), let us consider $n_0=1$ or $n'_0=1$. We have
\begin{equation}
W^{(d)}_{\Delta,\ell}=W^{(d-1)}_{\Delta,\ell}\;, \quad n_0=1\;,\;\;{\rm or}\;\; n'_0=1\;,
\end{equation}
which indicates these exchange Witten diagrams are {\it independent} of the spacetime dimension. This is easy to understand: the corresponding Mellin amplitudes has only the leading pole at $s=\Delta-\ell$, and its residue reproduces the collinear conformal block which is independent of $d$. 

Another observation is that a relation similar to (\ref{PSsymmetry}) also holds for the conformal partial waves \cite{Dobrev:1975ru,Dobrev:1977qv,SimmonsDuffin:2012uy}
\begin{equation}\label{CPWdef}
\Psi_{\Delta,\ell}^{(d)}=K^{34,(d)}_{\widetilde{\Delta},\ell}g^{(d)}_{\Delta,\ell}+K^{12,(d)}_{\Delta,\ell}g^{(d)}_{\widetilde{\Delta},\ell}
\end{equation}
where $\widetilde{\Delta}=d-\Delta$ is the conformal dimension for the shadow operator (we have left the $d$-dependence in the notation $\widetilde{\Delta}$ implicit), and 
\begin{equation}
K^{12,(d)}_{\Delta,\ell}=\left(-\frac{1}{2}\right)^\ell\frac{\pi^{\frac{d}{2}}\Gamma(\Delta-\frac{d}{2})\Gamma(\Delta+\ell-1)\Gamma(\frac{\widetilde{\Delta}+\Delta_1-\Delta_2+\ell}{2})\Gamma(\frac{\widetilde{\Delta}+\Delta_2-\Delta_1+\ell}{2})}{\Gamma(\Delta-1)\Gamma(d-\Delta+\ell)\Gamma(\frac{\Delta+\Delta_1-\Delta_2+\ell}{2})\Gamma(\frac{\Delta+\Delta_2-\Delta_1+\ell}{2})}\;.
\end{equation}
The new relation is obtained by simply replacing $g^{(d)}_{\Delta,\ell}$ in (\ref{PSsymmetry}) with $\Psi_{\Delta,\ell}^{(d)}$ and adjusting the $c_{i,j}$ coefficients accordingly to accommodate the extra $K$ factors
\begin{equation}
 \Psi_{\Delta,\ell}^{(d)}=\gamma_1\Psi^{(d)}_{\Delta,\ell}+\gamma_2 c_{2,0}\Psi^{(d)}_{\Delta+2,\ell}+\gamma_3 c_{1,-1}\Psi^{(d)}_{\Delta+1,\ell-1}+\gamma_4 c_{0,-2}\Psi^{(d)}_{\Delta,\ell-2}+\gamma_5 c_{2,-2}\Psi^{(d)}_{\Delta+2,\ell-2}
\end{equation} 
where 
\begin{equation}
\{\gamma_i\}=K^{34,(d-2)}_{\widetilde{\Delta},\ell}\big\{(K^{34,(d)}_{\widetilde{\Delta},\ell})^{-1},(K^{34,(d)}_{\widetilde{\Delta+2},\ell})^{-1},(K^{34,(d)}_{\widetilde{\Delta+1},\ell-1})^{-1},(K^{34,(d)}_{\widetilde{\Delta},\ell-2})^{-1},(K^{34,(d)}_{\widetilde{\Delta+2},\ell-2})^{-1}\big\}\;.
\end{equation}
 The shadow part will be satisfied automatically. Although verifying the relation is straightforward, we point out another interesting proof using the Witten diagram relation (\ref{Wreddtodm2}). We exploit the fact that the conformal partial waves can be represented as the difference of exchange Witten diagrams with opposite quantizations \cite{Hartman:2006dy,Costa:2014kfa,Giombi:2018vtc} ({\it i.e.}, with dimension $\Delta$ versus $d-\Delta$)\footnote{Here ${W'}_{\Delta,\ell}^{(d)}$ are defined by using the standard normalizations for the AdS propagators (the single-trace conformal block does not have coefficient one). The relation (\ref{PsiinWp}) can be easily proven by acting with the conformal Casimir on both sides. The r.h.s. solves the Casimir equation (\ref{Casimireqn}) thanks to the equation of motion relation (\ref{EOMid}) and the fact that contact terms do not depend on quantizations. Single-valuedness in Euclidean signature is also guaranteed by definition. See discussions around (2.47) in \cite{Zhou:2018sfz} for details of this proof.}
\begin{equation}\label{PsiinWp}
\Psi_{\Delta,\ell}^{(d)}\propto {W'}_{\Delta,\ell}^{(d)}-{W'}_{d-\Delta,\ell}^{(d)}\;.
\end{equation}
The new relation can be proven by taking the difference of (\ref{Wreddtodm2}) and  (\ref{Wreddtodm2}) with dimension $d-2-\Delta$ on the l.h.s. (the r.h.s. contains all the needed shadow diagrams in $AdS_{d+1}$).

Note that the arguments in this section which led to (\ref{Wreddtodm1}) and (\ref{Wreddtodm2}) assumed the existence of Mellin amplitudes. They therefore do not directly apply to the case of CFT$_1$, or CFT$_2$ with correlators not invariant under $z\leftrightarrow\bar{z}$, where the Mellin formalism is ill-defined\footnote{This is because the Mellin representation (\ref{defMellin}) assumed the cross ratios $U$ and $V$ to be independent. On the other hand, there is only one independent cross ratio in 1d. In 2d, the global conformal algebra factorizes into a left-moving and a right-moving part. The 2d CFT correlators are not necessarily symmetric in $z$ and $\bar{z}$. In such cases, one cannot unambiguously rewrite it in terms of $U$ and $V$, and the problem effectively becomes 1d. See \cite{Rastelli:2019gtj} for further discussions and examples of such mixed parity Witten diagrams in $AdS_3$.}. 
Nevertheless, the recursion relations should still be valid for these dimensions. First, the case of mixed parity CFT$_2$ correlators does not arise here because the conformal blocks in the recursion relations are all parity even. The corresponding $AdS_3$ Witten diagram should therefore also taken to be the type with even parity. Second, in reducing to 1d we should set $\ell=0$ since there is no spin in 1d (and also restrict to $z=\bar{z}$). The relations (\ref{Wreddtodm1}), (\ref{Wreddtodm2}) then only involve scalar exchange Witten diagrams. These diagrams can be evaluated analytically in $d$ (see Appendix C of \cite{Zhou:2018sfz} for explicit expressions). Therefore the Witten diagram relations above can also be analytically continued to hold in 1d. We will perform additional checks in  Appendix \ref{App:A} .

Finally, let us mention that the relations (\ref{Wreddtodm1}) and (\ref{Wreddtodm2}) can be verified in a number of explicit examples. The details of the calculations can be found in Appendix \ref{App:B}.

\section{Recursion relations involving differential operators}\label{Sec:3}
\subsection{The regular type}
In this section, we extend the strategy to include recursion relations with differential operators. To start, let us extract a kinematical factor 
\begin{equation}
G(x_i)=\frac{1}{(x_{12}^2)^{\frac{\Delta_1+\Delta_2}{2}}(x_{34}^2)^{\frac{\Delta_3+\Delta_4}{2}}}\left(\frac{x_{14}^2}{x_{24}^2}\right)^{\frac{\Delta_2-\Delta_1}{2}}\left(\frac{x_{14}^2}{x_{13}^2}\right)^{\frac{\Delta_3-\Delta_4}{2}}\mathcal{G}(U,V)\;,
\end{equation}
to express correlators in terms of the cross ratios. We further assume that the differential operators are of the form
\begin{equation}
\mathcal{D}=\sum_{\{m,n,a\}}\alpha_{m,n,a}\, U^mV^n\, \Omega_a(U\partial_U,V\partial_V)
\end{equation}
where $\Omega_a(U\partial_U,V\partial_V)$ are polynomials and the summation is over a finite set. In Mellin space, we can interpret the differential operator $\mathcal{D}$ as a difference operator $\widehat{\mathcal{D}}$ by interpreting the constituents as difference operators, which act on the Mellin amplitude $\mathcal{M}(s,t)$ in the following way
\begin{equation}
\begin{split}\label{basicdifference}
{}& \Omega_a(U\partial_U,V\partial_V)\;\mathcal{G}(U,V)\to\Omega_a(\tfrac{s}{2},\tfrac{t}{2}-\tfrac{\Delta_2+\Delta_3}{2})\times \mathcal{M}(s,t)\;,\\
{}& U^mV^n\;\mathcal{G}(U,V)\to \mathcal{M}(s-2m,t-2n)\left(\tfrac{\Delta_1+\Delta_2-s}{2}\right)_m\left(\tfrac{\Delta_3+\Delta_4-s}{2}\right)_m\left(\tfrac{\Delta_1+\Delta_4-t}{2}\right)_n\\
{}&\quad\quad\quad\quad\quad\quad\quad\quad\quad\quad\quad\quad\;\;\times \left(\tfrac{\Delta_2+\Delta_3-t}{2}\right)_n\left(\tfrac{s+t-\Delta_1-\Delta_3}{2}\right)_{-m-n}\left(\tfrac{s+t-\Delta_2-\Delta_4}{2}\right)_{-m-n}\;.
\end{split}
\end{equation}
This translation is clear from the definition of the Mellin amplitude (\ref{defMellin}).\footnote{A small subtlety is that the differential operators can change the integration contours in (\ref{defMellin}), and moving contours across poles may pick up extra terms (see the recent paper \cite{Penedones:2019tng} for a very detailed discussion on contours). We assume that it does not pose a problem for us, but we will also perform direct checks for the recursion relations in position space. Such issues do not seem to arise in the explicit examples in Appendix \ref{App:A} and Appendix \ref{App:B}.} To proceed, let us first focus on the simpler case when the difference operators $\widehat{\mathcal{D}}$ only shift the poles of the Mellin amplitude, and do not introduce new poles from the Pochhammer symbols that do not belong to the conformal blocks in the recursion relation.\footnote{These new poles are related to the double-trace operators. Physically, when we replace conformal blocks with exchange Witten diagrams we introduce additional double-trace conformal blocks. Under the action of the differential operators, some of the double-trace contributions may become singular.} We will call such recursion relations the {\it regular} type. It is clear that the singular parts of the Mellin amplitudes on both sides of the recursion relation are still matched, when we replace the conformal blocks with exchange Witten diagrams. However, the action of the difference operator may give rise to regular terms which cannot be absorbed into the redefinition of the exchange Witten diagrams. Nevertheless, such extra polynomial terms are easy to handle. They just correspond to additional contact Witten diagrams which need to be added into the Witten diagram recursion relations. 

Let us give two examples of such regular recursion relations, which involve the quadratic conformal Casimir operator
\begin{equation}
\begin{split}\label{defCasimir}
\mathrm{Cas}_s={}&2(UV^{-1}+1-V^{-1})V\frac{\partial}{\partial V}\left[V\frac{\partial}{\partial V}+a+b\right]-2U\frac{\partial}{\partial U}\left[2U\frac{\partial}{\partial U}-d\right]\\
{}&+2(1+U-V)\left[U\frac{\partial}{\partial U}+V\frac{\partial}{\partial V}+a\right]\left[U\frac{\partial}{\partial U}+V\frac{\partial}{\partial V}+b\right]
\end{split}
\end{equation}
where $a=\frac{\Delta_2-\Delta_1}{2}$, $b=\frac{\Delta_3-\Delta_4}{2}$. The first relation is the defining Casimir equation (\ref{Casimireqn}) for the s-channel conformal blocks $g^{(d)}_{\Delta,\ell}$. The second relation comes from acting the Casimir operator on the crossed channel conformal block, and it  leads to the following five-term recursion relation \cite{Zhou:2018sfz}
\begin{equation}\label{crossedCasimir}
\mathrm{Cas}_s[g^{(d),t}_{\Delta,\ell}]=A g^{(d),t}_{\Delta-1,\ell+1}+B g^{(d),t}_{\Delta-1,\ell-1}+C g^{(d),t}_{\Delta+1,\ell+1}+D g^{(d),t}_{\Delta+1,\ell-1}+E g^{(d),t}_{\Delta,\ell}\;.
\end{equation}
The coefficients are independent of the cross ratios, and are given in (\ref{ABCDE}) in Appendix \ref{App:B2}. This relation, for example, is useful for efficiently obtaining the conformal block decomposition coefficients of exchange Witten diagrams in the crossed channel \cite{Zhou:2018sfz}. From the operator (\ref{defCasimir}), one may worry that $(UV^{-1}+1-V^{-1})$ and $(1+U-V)$ can introduce new poles from the Pochhammer symbols in (\ref{basicdifference}). However, the operators $V\partial_V(V\partial_V+a+b)$ and $(U\partial_U+V\partial_V+a)(U\partial_U+V\partial_V+b)$ introduce polynomial factors which precisely cancel these poles. The whole operator in Mellin space therefore only reshuffles pole locations, but does not introduce new singularities. As a result, when we replace the conformal blocks with exchange Witten diagrams, all the polar terms of the Mellin amplitudes come from the conformal blocks. They are guaranteed to match because their residues simply reproduce the conformal block recursion relations (\ref{Casimireqn}) and (\ref{crossedCasimir}). We yet need to match the regular terms in Mellin amplitudes. It turns out that the Casimir operator raises the degree of  polynomials only by one, despite that it contains two derivatives.   Therefore, from (\ref{EOMid}) we obtain the following recursion relation for Witten diagrams
\begin{equation}
\mathrm{Cas}_s[W^{(d)}_{\Delta,\ell}]-\mathcal{C}_{\Delta,\ell}\,W^{(d)}_{\Delta,\ell}=W^{(d)}_{con}\;,
\end{equation}
and from (\ref{crossedCasimir}) we have
\begin{equation}\label{crossedCasimirforW}
\mathrm{Cas}_s[W^{(d),t}_{\Delta,\ell}]=A W^{(d),t}_{\Delta-1,\ell+1}+B W^{(d),t}_{\Delta-1,\ell-1}+C W^{(d),t}_{\Delta+1,\ell+1}+D W^{(d),t}_{\Delta+1,\ell-1}+E W^{(d),t}_{\Delta,\ell}\;.
\end{equation}
Note that in the first relation, a contact Witten diagram $W^{(d)}_{con}$ with $2\ell$ derivatives needs to be explicitly added, while in the second case it can be absorbed into $W^{(d),t}_{\Delta+1,\ell+1}$. Several explicit examples of the relation  (\ref{crossedCasimirforW}) are given in Appendix \ref{App:B}.

\subsection{The irregular type}
Now let us comment on what happens when the differential operator introduces new poles in Mellin space (we will refer to it as the {\it irregular} type). To make the point, it is sufficient to focus on the following example of recursion relation \cite{Dolan:2011dv}
\begin{equation}\label{reF0}
(U^{-1}-U^{-1}V)g^{(d)}_{\Delta,\ell}=A' g^{(d)}_{\Delta-1,\ell+1}+B' g^{(d)}_{\Delta-1,\ell-1}+C' g^{(d)}_{\Delta+1,\ell+1}+D' g^{(d)}_{\Delta+1,\ell-1}+E' g^{(d)}_{\Delta,\ell}\;.
\end{equation}
The coefficients can be found in (4.32) of \cite{Dolan:2011dv}, however their explicit forms are not important for the following discussion. Let us first attempt to get a relation for Witten diagrams by naively replacing the conformal blocks with the exchange Witten diagrams, and translating the cross ratio dependent factor into a difference operator. However, we immediately encounter a problem. In Mellin space, we find the factor $(U^{-1}-U^{-1}V)$ acts on the Mellin amplitude as 
\begin{equation}\label{actionF0}
\frac{-(t-\Delta_2-\Delta_3)(t-\Delta_1-\Delta_4)\mathcal{M}(s+2,t-2)+(u-\Delta_2-\Delta_4)(u-\Delta_1-\Delta_3)\mathcal{M}(s+2,t))}{(s-(\Delta_1+\Delta_2-2))(s-(\Delta_3+\Delta_4-2))}\;.
\end{equation}
The $s\to s+2$ shift preserves the series of poles at $s=\Delta-\ell+2m$, and can be accounted for by the conformal blocks on the r.h.s. of (\ref{reF0}). The poles at $s=\Delta_1+\Delta_2-2$ and $s=\Delta_3+\Delta_4-2$ however are new. To match these singularities, we must add extra Witten diagrams. These additional terms are the linear combination of $(U^{-1}-U^{-1}V)W^{(d)}_{con,1}$, $W^{(d)}_{\Delta_1+\Delta_2-2+\ell',\ell'}$, $W^{(d)}_{\Delta_3+\Delta_4-2+\ell',\ell'}$ and $W^{(d)}_{con,2}$. Their analytic structures in Mellin space makes it clear that they are sufficient to cancel any remaining terms.\footnote{The first kind of diagrams are needed because (\ref{actionF0}) gives rise simultaneous poles in $s$, while the exchange Witten diagrams have only single poles with polynomial residues.} In Appendix \ref{App:B}, we will study examples with $\ell=0$ both in Mellin and in position space, and obtain explicit expressions for such extra terms. 

Many other interesting relations fall into this category. For example, there are various generalizations of (\ref{reF0}) in \cite{Dolan:2011dv} relating conformal blocks with shifted dimensions and spins, which can be systematized using the weight-shifting operators \cite{Karateev:2017jgd}. Moreover, when supersymmetry is present there are further kinematic constraints in the form of superconformal Ward identities. These identities give rise to nontrivial recursion relations for the non-supersymmetric conformal blocks in the superconformal block.  The superconformal blocks for  short multiplets are particularly interesting, as they are relevant for the holographic calculation of boundary correlators in the supergravity limit (see Appendix \ref{App:B} for more comments). The exchange of short superconformal blocks can be identified with the  supergravity  exchange diagrams in the bulk, modulo double-trace operators.  Since these relations are in general of the irregular type,  extra terms are needed when translating them into exchange Witten diagrams as illustrated above.

\section{Generalization to CFTs with boundaries}\label{Sec:4}
In this section we generalize the story to include boundary CFTs with a flat conformal boundary (or co-dimension 1 interface). We break the coordinates of $\mathbb{R}^d$ into $x^\mu=(x_\perp, \vec{x})$, where $x_\perp$ and $\vec{x}$ are the directions transverse and parallel to the boundary respectively. The boundary is at $x_\perp=0$. We refer the reader to, {\it e.g.}, \cite{Liendo:2012hy} for a detailed account about the basic kinematics, and will be brief in the following. 

We focus on the simplest correlators with nontrivial spacetime dependence, namely, two-point functions   
\begin{equation}
\langle \mathcal{O}_1(x_1)\mathcal{O}_2(x_2)\rangle=\frac{\mathcal{G}(\xi)}{(2x_{1,\perp})^{\Delta_1}(2x_{2,\perp})^{\Delta_2}}
\end{equation}
where the cross ratio $\xi$ is defined as 
\begin{equation}
\xi=\frac{(x_1-x_2)^2}{4x_{1,\perp}x_{2,\perp}}\;,
\end{equation}
and $(x_1-x_2)=(x_{1,\perp}-x_{2,\perp})^2+(\vec{x}_1-\vec{x}_2)^2$. The correlator $\mathcal{G}(\xi)$ can be decomposed into conformal blocks in two channels
\begin{equation}
\mathcal{G}(\xi)=\sum_k \mu_{12k}\, g^{(d),bulk}_{\Delta_k}(\xi)=\sum_j \hat{\mu}_{12j}\, g^{(d),bdry}_{\Delta_j}(\xi)
\end{equation}
where in the bulk channel $\mathcal{O}_1$, $\mathcal{O}_2$ are merged to form one-point functions, and in the boundary channel $\mathcal{O}_1$, $\mathcal{O}_2$ approach the boundary and form two-point functions on the boundary. The conformal blocks are given by \cite{McAvity:1995zd,Liendo:2012hy}
\begin{equation}\label{defgbulk}
g^{(d),bulk}_{\Delta}(\xi)=\xi^{\frac{\Delta-\Delta_1-\Delta_2}{2}} {}_2F_1\left(\frac{\Delta+\Delta_1-\Delta_2}{2},\frac{\Delta+\Delta_2-\Delta_1}{2};\Delta-\frac{d}{2}+1;-\xi\right)\;,
\end{equation}
\begin{equation}\label{defgboundary}
g^{(d),bdry}_{\Delta}(\xi)=\xi^{-\Delta}{}_2F_1\left(\Delta,\Delta-\frac{d}{2}+1;2\Delta+2-d,-\frac{1}{\xi}\right)\;.
\end{equation}
Using these explicit expressions, it is not difficult to verify the following recursion relations.  For reducing from $d$ to $d-1$ dimensions, we have the formulae\footnote{It is also possible to derive formulae that go in the opposite directions
\begin{equation}\label{dtodp1bulk}
g^{(d),bulk}_{\Delta}(\xi)=\sum_{n=0}^\infty \frac{d-2 \Delta -4 n+1}{(1-2 n) (d-2 \Delta -2 n+1)}\beta_n\, g^{(d+1),bulk}_{\Delta+2n}(\xi)\;,
\end{equation}
\begin{equation}\label{dtodp1boundary}
g^{(d),bdry}_{\Delta}(\xi)=\sum_{n=0}^\infty  \frac{d-2 \Delta -4 n}{(1-2 n) (d-2 \Delta -2 n)}  \gamma_n\, g^{(d+1),bdry}_{\Delta+2n}(\xi)\;.
\end{equation}}
\begin{equation}\label{gbulkdtodm1}
g^{(d),bulk}_{\Delta}(\xi)=\sum_{n=0}^\infty \beta_n\, g^{(d-1),bulk}_{\Delta+2n}(\xi)\;,
\end{equation}
\begin{equation}\label{gboundarydtodm1}
g^{(d),bdry}_{\Delta}(\xi)=\sum_{n=0}^\infty \gamma_n\, g^{(d-1),bdry}_{\Delta+2n}(\xi)
\end{equation}
where the coefficients are 
\begin{equation}
\beta_n=\frac{(-1)^n \Gamma \left(n+\frac{1}{2}\right) \left(\frac{\Delta +\Delta_1-\Delta_2}{2}\right)_n \left(\frac{\Delta -\Delta_1+\Delta_2}{2}\right)_n}{\sqrt{\pi } n! \left(-\frac{d}{2}+\Delta +1\right)_n \left(-\frac{d}{2}+n+\Delta +\frac{1}{2}\right)_n}\;,
\end{equation}
\begin{equation}
\gamma_n=\frac{2^{-4 n} \Gamma \left(n+\frac{1}{2}\right) (\Delta )_{2 n}}{\sqrt{\pi } \Gamma (n+1) \left(-\frac{d}{2}+\Delta +\frac{3}{2}\right)_n \left(-\frac{d}{2}+n+\Delta +1\right)_n}\;.
\end{equation}
On the other hand, we find that the $d-2$ dimensional conformal blocks can be expressed in terms of just two $d$ dimensional conformal blocks\footnote{By contrast, expressing $d$ dimensional conformal blocks in terms the $d-2$ dimensional ones requires infinitely many terms
\begin{equation}\label{dtodm2bulk}
g^{(d),bulk}_{\Delta}(\xi)=\sum_{n=0}^\infty \tfrac{(-1)^n \left(\left(\frac{1}{2} (\Delta -\Delta_1+\Delta_2)\right)_n \left(\frac{1}{2} (\Delta +\Delta_1-\Delta_2)\right)_n\right)}{\left(-\frac{d}{2}+n+\Delta +1\right)_n \left(-\frac{d}{2}+\Delta +1\right)_n}\, g^{(d-2),bulk}_{\Delta+2n}(\xi)\;,
\end{equation}
\begin{equation}\label{dtodm2boundary}
g^{(d),bdry}_{\Delta}(\xi)=\sum_{n=0}^\infty \tfrac{16^{-n} (\Delta )_{2 n}}{\left(-\frac{d}{2}+\Delta +\frac{3}{2}\right)_{2 n}}\, g^{(d-2),bdry}_{\Delta+2n}(\xi)\;.
\end{equation}}
\begin{equation}\label{gbulkdtodm2}
g^{(d-2),bulk}_{\Delta}(\xi)=g^{(d),bulk}_{\Delta}(\xi)+\left(\tfrac{(\Delta +\Delta_1-\Delta_2) (\Delta -\Delta_1+\Delta_2)}{(d-2 \Delta -4) (d-2 \Delta -2)}\right)g^{(d),bulk}_{\Delta+2}(\xi)\;,
\end{equation}
\begin{equation}\label{gboundarydtodm2}
g^{(d-2),bdry}_{\Delta}(\xi)=g^{(d),bdry}_{\Delta}(\xi)-\left(\tfrac{\Delta  (\Delta +1)}{4 (d-2 \Delta -5) (d-2 \Delta -3)}\right)g^{(d),bdry}_{\Delta+2}(\xi).
\end{equation}
The existence of such two-term recursion relations is quite remarkable, and seems to suggest a generalization of the Parisi-Sourlas supersymmetry to include boundaries. 

As a side comment, let us mention that the boundary channel conformal block in $d$ dimensions is identical to the scalar bulk-to-bulk propagator $G_{BB}^{(d-1),\Delta}(u)$ in $AdS_d$\footnote{This fact was noticed in \cite{Rastelli:2017ecj}, and was used to find the geodesic Witten diagram representation for the boundary channel conformal block.}
\begin{equation}
G_{BB}^{(d-1),\Delta}(u)=\frac{\pi ^{\frac{1-d}{2}} (-4)^{-\Delta } \Gamma (\Delta )}{2\Gamma \left(-\frac{d}{2}+\Delta +\frac{3}{2}\right)}\, g^{(d),bdry}_{\Delta}(u)
\end{equation}
where $u=\frac{(z-w)^2}{2z_0w_0}$ is the chordal distance\footnote{Here $z=(z_0,\vec{z})$ are the Poincar\'e coordinates.} 
 between $z$ and $w$ in $AdS_d$, and the propagator satisfies the equation of motion 
 \begin{equation}
( \square-\Delta(\Delta-d+1))G_{BB}^{(d-1),\Delta}=\delta(z,w)\;.
 \end{equation}
Thanks to this identification, (\ref{gbulkdtodm1}), (\ref{gboundarydtodm1}), (\ref{gbulkdtodm2}) and (\ref{gboundarydtodm2}) can also be interpreted as the dimensional reduction formulae for the AdS bulk-to-bulk propagator. Using these relations, it is straightforward to use geodesic Witten diagrams \cite{Hijano:2015zsa} to prove the conformal block recursion relations (\ref{dtodm1}) and (\ref{PSsymmetry}) for the scalar case.

The above reduction formulae for the conformal blocks also imply recursion relations for exchange Witten diagrams in the so-called probe brane setup for interface CFTs (the simplest version of the Karch-Randall setup \cite{Karch:2001cw,Karch:2000gx}). In this setup we single out an $AdS_d$ slice inside of $AdS_{d+1}$, which is located at $z_\perp=0$ in the Poincar\'e coordinates. The conformal boundary of the $AdS_d$ slice is the interface of the CFT on the boundary. We allow local degrees of freedom on the $AdS_d$ probe brane, and they can be coupled to the bulk $AdS_{d+1}$ fields. However the probe does not back-react to the bulk geometry. One can define the following tree level exchange Witten diagrams which describe perturbative interactions in the effective theory
\begin{equation}
W^{(d),bulk}_\Delta=N^{(d)}_{bulk}\int_{AdS_d}d^dw\int_{AdS_{d+1}}d^{d+1}z\; G_{BB}^{(d),\Delta}(z,w)G^{(d),\Delta_1}_{B\partial}(x_1,z)G^{(d),\Delta_2}_{B\partial}(x_2,z)\;,
\end{equation}
\begin{equation}
W^{(d),bdry}_\Delta=N^{(d)}_{dbry}\int_{AdS_d}d^dw_1d^dw_2\;G_{BB}^{(d-1),\Delta}(w_1,w_2)G^{(d),\Delta_1}_{B\partial}(x_1,w_1)G^{(d),\Delta_2}_{B\partial}(x_2,w_2)\;,
\end{equation}
where $G^{(d),\Delta_2}_{B\partial}(x,z)$ are the bulk-to-boundary propagators in $AdS_{d+1}$. We have also inserted normalization factors $N^{(d)}_{bulk}$, $N^{(d)}_{dbry}$ such that the single-trace conformal blocks appear with coefficient one in the direct channel. These diagrams have been systematically studied in \cite{Rastelli:2017ecj,Mazac:2018biw}, and will not be further commented on here. The only ingredient we want to highlight is that  interface CFT correlators also admit a Mellin representation \cite{Rastelli:2017ecj}. The two-point exchange Mellin amplitudes are functions of a single Mellin-Mandelstam variable, and has only simple poles corresponding to exchanged single-trace operator, with constant residues (see \cite{Rastelli:2017ecj} for details). This allows us to use the Mellin argument in Section \ref{Sec:2}, and write down the following recursion relations for the exchange Witten diagrams 
\begin{eqnarray}
&&W^{(d),bulk}_{\Delta}(\xi)=\sum_{n=0}^\infty \beta_n\, W^{(d-1),bulk}_{\Delta+2n}(\xi)\;,\quad W^{(d),bdry}_{\Delta}(\xi)=\sum_{n=0}^\infty \gamma_n\, W^{(d-1),bdry}_{\Delta+2n}(\xi)\;,\\
&&W^{(d-2),bulk}_{\Delta}(\xi)=W^{(d),bulk}_{\Delta}(\xi)+\left(\tfrac{(\Delta +\Delta_1-\Delta_2) (\Delta -\Delta_1+\Delta_2)}{(d-2 \Delta -4) (d-2 \Delta -2)}\right)W^{(d),bulk}_{\Delta+2}(\xi)\;,\\
&&W^{(d-2),bdry}_{\Delta}(\xi)=W^{(d),bdry}_{\Delta}(\xi)-\left(\tfrac{\Delta  (\Delta +1)}{4 (d-2 \Delta -5) (d-2 \Delta -3)}\right)W^{(d),bdry}_{\Delta+2}(\xi),
\end{eqnarray}
and similarly the counterparts for (\ref{dtodp1bulk}), (\ref{dtodp1boundary}), (\ref{dtodm2bulk}) and (\ref{dtodm2boundary}). We can also consider more complicated Witten diagram relations which descend from recursion relations of conformal blocks involving differential operators. However the logic is largely similar, and we will not repeat the analysis here.

\section{Discussion and outlook}\label{Sec:5}
In this paper, we pointed out that conformal block recursion relations in CFT naturally give rise to recursion relations for exchange Witten diagrams in AdS. This opens the door to a wealth of new properties about Witten diagrams which are difficult to discover otherwise. Note that our statement is different from replacing conformal blocks with geodesic Witten diagrams \cite{Hijano:2015zsa}, because the latter is just representing the same functions in two different ways and does not generate new identities. By contrast, exchange Witten diagrams contain in addition infinitely many double-trace conformal blocks. The cancellation of the double-trace operators in the identities are highly nontrivial. In the paper we presented concrete methods for obtaining these Witten diagram relations, and applied the methods to many examples. We derived a variety of useful identities for studying AdS scattering, including the dimensional reduction formulae for exchange Witten diagrams. These examples by no means have exhausted all the applications, and it would be interesting to find more relations. We also outlined the generalization to CFTs with boundaries. We gave the dimensional reduction formulae for BCFT conformal blocks, and also wrote down the corresponding relations for exchange Witten diagrams in the probe brane setup.

Our work leads to many avenues for future research. 

One interesting extension is to consider recursion relations of conformal blocks for spinning correlators. By extending the logic of this paper, these relations should imply recursion relations for exchange Witten diagrams with spinning external states. They will be useful for studying, for example, properties of gauge theory scattering amplitudes in AdS. In fact, there are already some hints that such results might be possible. On the CFT side, there are weight-shifting operators which change the representations of the operators \cite{Karateev:2017jgd}. On the AdS side, there are also analogous operators which act on AdS harmonics and bulk-to-boundary propagators \cite{Costa:2018mcg}. Using these weight-shifting operators, one can reduce spinning objects to just the scalar ones on both sides. To firmly establish the claim, it is perhaps best to work with the Mellin representation for spinning correlators \cite{Goncalves:2014rfa,Chen:2017xdz,Sleight:2018epi,Binder:2020raz}. The arguments in this paper can then be adapted to argue that the double-trace operators will cancel out.   

Another direction is to explore generalizations to higher-point correlators. There have been some recent progress in studying conformal blocks with five or more external operators \cite{Rosenhaus:2018zqn, Parikh:2019ygo,Goncalves:2019znr,Jepsen:2019svc,Parikh:2019dvm,Fortin:2019zkm,Fortin:2020yjz}. It would be interesting to search for recursion relations and identify their Witten diagram counterparts. 

Exchange Witten diagrams are also intimately related to the method of analytic functional bootstrap \cite{Mazac:2016qev,Mazac:2018mdx,Mazac:2018ycv,Mazac:2018biw,Kaviraj:2018tfd,Hartman:2019pcd,Paulos:2019gtx,Mazac:2019shk,Carmi:2019cub,Huang:2019xzm,Sleight:2019ive} (see also \cite{Gopakumar:2016wkt,Gopakumar:2016cpb,Gopakumar:2018xqi,Ferrero:2019luz}), as they neatly encapsulate the information of the functionals. More precisely, the Polyakov-Regge blocks can be decomposed into conformal blocks in two channels. The double-trace coefficients are identified (up to signs) with the action of a basis of analytic functionals on the single-trace conformal block \cite{Mazac:2018ycv,Mazac:2018biw,Mazac:2019shk}. Using this property together with the Witten diagram recursion relations, we now obtain a zoo of new relations for the analytic functionals. To see this, let us decompose both sides of the Witten diagram recursion relations into conformal blocks (either in the direct channel or in the crossed channel). This can be achieved by first inserting the conformal block decomposition of the Polyakov-Regge blocks, and using the  recursion relations for conformal blocks. The matching of the double-trace conformal blocks then imply nontrivial recursion relations for the functional actions (an example is given in \ref{App:A}). These relations might provide insights into new structures in the analytic functionals. One can also ask if these relations can aid the calculation of extracting functional actions from Witten diagrams in higher dimensions \cite{Mazac:2019shk}. We will leave these questions for future work.

Finally, we noticed in Section \ref{Sec:4} that BCFT conformal blocks in $d$ dimensions can be expressed in terms of just two conformal blocks in $d-2$ dimensions (see (\ref{gbulkdtodm2}) and (\ref{gboundarydtodm2})). On the other hand, reversing the relation would require infinitely many conformal blocks.  This is quite reminiscent of the situation in \cite{Kaviraj:2019tbg}, and provides kinematical evidence for the generalization of the Parisi-Sourlas supersymmetry \cite{Parisi:1979ka} to include boundaries. It would be interesting to explore this idea in detail.

\acknowledgments
This work is supported in part by Simons Foundation Grant No. 488653.

\appendix

\section{Scalar diagrams and reduction to one dimension}\label{App:A}
In this Appendix, we perform some checks for reductions involving only scalars. This is related to the reduction to 1d where there is no spin. Setting $\ell=0$, the recursion relations for conformal blocks read
\begin{equation}\label{greddtodm1sca}
g^{(d)}_{\Delta,0}=\sum_{n=0}^{\infty}A_{n,0}\, g^{(d-1)}_{\Delta+2n,0}\;,
\end{equation}
\begin{equation}\label{greddtodm2sca}
g^{(d-2)}_{\Delta,0}=g^{(d)}_{\Delta,0}+c_{2,0}g^{(d)}_{\Delta+2,0}\;,
\end{equation}
and correspondingly,
\begin{equation}\label{Wreddtodm1sca}
W^{(d)}_{\Delta,0}=\sum_{n=0}^{\infty}A_{n,0}\, W^{(d-1)}_{\Delta+2n,0}\;,
\end{equation}
\begin{equation}\label{Wreddtodm2sca}
W^{(d-2)}_{\Delta,0}=W^{(d)}_{\Delta,0}+c_{2,0}W^{(d)}_{\Delta+2,0}\;.
\end{equation}

We will perform the checks in position space, and proceed by decomposing each scalar exchange Witten diagram into conformal blocks.  The decomposition coefficients can be written down for general $d$. Therefore we will keep $d$ arbitrary and treat the reduction to 1d as a special example. We will also keep $\Delta_i$ general in order to avoid complications of  associated with overlapping double-trace spectra.\footnote{In the coinciding limit of the double-trace spectra, derivatives of the conformal blocks $\partial_\Delta g^{(d)}_{\Delta,0}$ arise and compensate half of the double-trace conformal blocks which have now become degenerate.} The coefficients $A_{n,0}$ are given by
\begin{equation}
A_{n,0}=\frac{(-1)^n \left(\frac{3}{2}\right)_{n-1} \left(\frac{\Delta +\Delta_1-\Delta_2}{2}\right)_n \left(\frac{\Delta -\Delta_1+\Delta_2}{2}\right)_n \left(\frac{\Delta +\Delta_3-\Delta_4}{2}\right)_n \left(\frac{\Delta -\Delta_3+\Delta_4}{2}\right)_n}{2 (2)_{n-1} (\Delta )_{2 n} \left(\frac{d-4 n-2 \Delta +1}{2}\right)_n \left(-\frac{d}{2}+\Delta +1\right)_n}\;,
\end{equation}
and  reproduce (\ref{Anj}) upon setting $\Delta_i$.  For the exchange Witten diagrams we have the following decomposition
\begin{equation}
W^{(d)}_{\Delta,0}=g^{(d)}_{\Delta,0}+\sum_{n=0}^\infty \mu^{(d)}_{12,n}(\Delta) g^{(d)}_{\Delta^{12}_n,0}+\sum_{n=0}^\infty \mu^{(d)}_{34,n}(\Delta) g^{(d)}_{\Delta^{34}_n,0}\;,
\end{equation}
where $\Delta^{12}_n=\Delta_1+\Delta_2+2n$, $\Delta^{34}_n=\Delta_3+\Delta_4+2n$, and the coefficients are given by (see, {\it e.g.}, \cite{Zhou:2018sfz})
\begin{equation}
\begin{split}
\mu^{(d)}_{12,n}(\Delta)=\frac{\nu^{(d)}_{12,n}(\Delta)}{\rho^{(d)}(\Delta)}\;,
\end{split}
\end{equation}
with 
\begin{equation}
\begin{split}
\nu^{(d)}_{12,n}={}&\frac{ (-1)^{n} \Gamma \left(\frac{-2 n-\Delta_1-\Delta_2+\Delta_3+\Delta_4}{2}\right) \Gamma \left(\frac{2 n+\Delta_1+\Delta_2+\Delta_3-\Delta_4}{2}\right) \Gamma \left(\frac{2 n+\Delta_1+\Delta_2-\Delta_3+\Delta_4}{2}\right)}{n! \Gamma (2 n+\Delta_1+\Delta_2)\Gamma \left(-\frac{d}{2}+2 n+\Delta_1+\Delta_2\right)}\\
{}&\times\frac{ \Gamma (n+\Delta_1) \Gamma (n+\Delta_2)\Gamma \left(-\frac{d}{2}+n+\Delta_1+\Delta_2\right)  \Gamma \left(\frac{-d+2 n+\Delta_1+\Delta_2+\Delta_3+\Delta_4}{2}\right)}{ (-\Delta +\Delta_1+\Delta_2+2 n) (d-\Delta -\Delta_1-\Delta_2-2 n) }\;,
\end{split}
\end{equation}
and
\begin{equation}
\begin{split}
\rho^{(d)}={}&\frac{ \Gamma \left(\frac{\Delta +\Delta_1-\Delta_2}{2}\right) \Gamma \left(\frac{\Delta -\Delta_1+\Delta_2}{2}\right) \Gamma \left(\frac{-\Delta +\Delta_1+\Delta_2}{2}\right) \Gamma \left(\frac{\Delta +\Delta_3-\Delta_4}{2}\right) \Gamma \left(\frac{\Delta -\Delta_3+\Delta_4}{2}\right)  }{4 \Gamma (\Delta ) \Gamma \left(-\frac{d}{2}+\Delta +1\right)}\\
{}&\times\Gamma \left(\tfrac{-d+\Delta +\Delta_1+\Delta_2}{2}\right) \Gamma \left(\tfrac{-d+\Delta +\Delta_3+\Delta_4}{2}\right)\Gamma \left(\tfrac{-\Delta +\Delta_3+\Delta_4}{2} \right)\;.
\end{split}
\end{equation}
The coefficients $\mu^{(d)}_{34,n}$ can be obtained from $\mu^{(d)}_{12,n}$ from replacing $\Delta_1$, $\Delta_2$ with $\Delta_3$, $\Delta_4$. The single-trace conformal blocks are cancelled from the relation (\ref{Wreddtodm2sca}), and we have the following constraint on the double-trace coefficients 
\begin{equation}
\sum_{n=0}^\infty \mu^{(d-2)}_{12,n}(\Delta)g^{(d-2)}_{\Delta^{12}_n,0}=\sum_{n=0}^\infty \mu^{(d)}_{12,n}(\Delta)g^{(d)}_{\Delta^{12}_n,0}+c_{2,0}(\Delta)\sum_{n=0}^\infty \mu^{(d)}_{12,n}(\Delta+2)g^{(d)}_{\Delta^{12}_{n},0}\;,
\end{equation} 
and similarly for double-trace operators made of 3 and 4. Using (\ref{greddtodm2sca}), we can rewrite it in terms conformal blocks in $d$ dimensions
\begin{equation}
\sum_{n=0}^\infty \mu^{(d-2)}_{12,n}(\Delta) (g^{(d)}_{\Delta^{12}_n,0}+c_{2,0}(\Delta^{12}_n)g^{(d)}_{\Delta^{12}_{n+1},0})=\sum_{n=0}^\infty (\mu^{(d)}_{12,n}(\Delta)+c_{2,0}(\Delta) \mu^{(d)}_{12,n}(\Delta+2))g^{(d)}_{\Delta^{12}_n,0}\;.
\end{equation} 
It is straightforward to check 
\begin{equation}\label{DTcoefre}
\mu^{(d-2)}_{12,n}(\Delta)+c_{2,0}(\Delta^{12}_{n-1})\mu^{(d-2)}_{12,n-1}(\Delta)=\mu^{(d)}_{12,n}(\Delta)+c_{2,0}(\Delta) \mu^{(d)}_{12,n}(\Delta+2)\;,
\end{equation}
and the above condition is satisfied (we define $\mu^{(d-2)}_{12,-1}(\Delta)=0$). The condition on the $g^{(d)}_{\Delta^{34}_{n},0}$ can be similarly verified. Let us point out that the decomposition coefficients $\mu^{(d)}_{12,n}(\Delta)$ can be viewed as the action of functionals on the scalar conformal block $g^{(d)}_{\Delta,0}$ (see \cite{Mazac:2019shk} for details). The above recursion relation therefore yields relations for the analytic functionals. We can also analyze the conformal block decomposition of (\ref{Wreddtodm2sca}) in the t-channel, which gives relations for the actions of the t-channel functionals.

Checking (\ref{Wreddtodm1sca}) is similar to the finite term case, however we need to perform infinite summations for each double-trace conformal block coefficient. Although it is quite difficult to explicitly perform the infinite sum for generic $\Delta_i$ and $\Delta$, one can numerically convince oneself that the relation (\ref{Wreddtodm1sca}) is valid.

\section{Explicit examples}\label{App:B}
\subsection{Dimensional reduction}
Let us verify the reduction formulae (\ref{Wreddtodm1}) and (\ref{Wreddtodm2}) in a number of explicit examples. We start again with $\ell=0$ on the l.h.s., but now in Mellin space. The relations can be easily verified with the explicit formula for the Mellin amplitude\footnote{The coefficients $a^{(d)}_m$ can be easily obtained from solving the recursion relations induced by the equation of motion identity (\ref{EOMid}) in Mellin space. The vector exchange Mellin amplitude is obtained similarly.} 
\begin{equation}\label{Mellinscalar}
\mathcal{M}^{(d)}_{\Delta,0}(s,t)=\sum_{m=0}^\infty \frac{a^{(d)}_m}{s-\Delta-2m}
\end{equation}
where 
\begin{equation}
\begin{split}
a^{(d)}_m={}& \frac{1}{\Gamma \left(\frac{\Delta +\Delta_1-\Delta_2}{2}\right) \Gamma \left(\frac{\Delta -\Delta_1+\Delta_2}{2}\right) \Gamma \left(\frac{-\Delta +\Delta_1+\Delta_2}{2} \right)}\\
\times{}&\frac{2 \Gamma (\Delta ) \left(\frac{\Delta -\Delta_1-\Delta_2+2}{2}\right)_m \left(\frac{\Delta -\Delta_3-\Delta_4+2}{2}\right)_m}{m! \left(-\frac{d}{2}+\Delta +1\right)_m  \Gamma \left(\frac{\Delta +\Delta_3-\Delta_4}{2}\right) \Gamma \left(\frac{\Delta -\Delta_3+\Delta_4}{2}\right) \Gamma \left(\frac{-\Delta +\Delta_3+\Delta_4}{2}\right)}\;.
\end{split}
\end{equation}
Let us now move on to $\ell=1$. For simplicity, we restrict to  pairwise equal conformal dimensions $\Delta_1=\Delta_2=p$, $\Delta_3=\Delta_4=q$. The s-channel exchange Mellin amplitude has the following form\footnote{Here we have chosen $P^{(d)}_{0}=0$ for simplicity.} 
\begin{equation}\label{Mellinvec1}
\mathcal{M}^{(d)}_{\Delta,1}(s,t)=\sum_{m=0}^\infty\frac{b^{(d)}_m\big(2(p+q-t)-(\Delta-1+2m)\big)}{s-(\Delta-1)-2m}
\end{equation}
where 
\begin{equation}
b^{(d)}_m=\frac{4 \Delta  \Gamma (\Delta -1) \left(\frac{1}{2} (-2 p+\Delta +1)\right)_m \left(\frac{1}{2} (-2 q+\Delta +1)\right)_m}{(\Delta -1)^2 \Gamma \left(\frac{\Delta -1}{2}\right)^4 \Gamma (m+1) \Gamma \left(p-\frac{\Delta }{2}+\frac{1}{2}\right) \Gamma \left(q-\frac{\Delta }{2}+\frac{1}{2}\right) \left(-\frac{d}{2}+\Delta +1\right)_m}\;.
\end{equation}
It is straightforward to see that (\ref{Wreddtodm1}) and  (\ref{Wreddtodm2}) are satisfied. 

We can also verify the relations in position space.  Let us illustrate this with the case where $\Delta_i=\Delta=3$, $\ell=1$. With these quantum numbers, we can use the method of \cite{DHoker:1999mqo} and the exchange Witten diagrams truncate and can be expressed as a finite sum of $\bar{D}$-functions. Using the formulae in \cite{DHoker:1999mqo}, we get\footnote{Note that the choice of the cubic vertices in \cite{DHoker:1999mqo} leads to a different choice with non-vanishing $P^{(d)}_{\Delta,\ell}$. We denote them with an extra tilde to distinguish it from the other choice made above.}
\begin{equation}\label{W31}
\begin{split}
\tilde{W}^{(d)}_{3,1}={}&-\frac{3U}{2x_{12}^6x_{34}^6}\bigg((\bar{D}_{1234}-V\bar{D}_{1243}-\bar{D}_{2134}+\bar{D}_{2143})\\
{}&+\frac{d-10}{d-8}U(\bar{D}_{2334}-V\bar{D}_{2343}-\bar{D}_{3234}+\bar{D}_{3243})\bigg)\;,
\end{split}
\end{equation}
\begin{equation}\label{W51}
\tilde{W}^{(d)}_{5,1}=-\frac{15U^2}{4x_{12}^6x_{34}^6}(\bar{D}_{2334}-V\bar{D}_{2343}-\bar{D}_{3234}+\bar{D}_{3243})\;.
\end{equation}
It is straightforward to find  
\begin{equation}
\tilde{W}^{(d)}_{3,1}=\tilde{W}^{(d-1)}_{3,1}+\frac{4}{5 (d-9) (d-8)}\tilde{W}^{(d-1)}_{5,1}\;,
\end{equation}
\begin{equation}
\tilde{W}^{(d-2)}_{3,1}=\tilde{W}^{(d)}_{3,1}-\frac{8}{5 (d-10) (d-8)}\tilde{W}^{(d)}_{5,1}\;,
\end{equation}
which reproduce correspondingly the special cases of (\ref{Wreddtodm1}) and (\ref{Wreddtodm2}).

\subsection{Recursion relations with differential operators}\label{App:B2}
In this subsection, we give a few explicit examples of Witten diagram recursion relations involving differential operators. 

\vspace{0.3cm}
\noindent{\bf The regular type}
\vspace{0.1cm}

\noindent To start, let us give the coefficients in (\ref{crossedCasimir}). We will focus on the case where $\Delta_i=\Delta_\phi$, and the coefficients read
\begin{eqnarray}
\nonumber {}&&A= (-\Delta +2 \Delta_\phi +\ell )^2\;,\\
\nonumber {}&&B=\frac{\ell  (d+\ell -3) (d+\Delta -2 \Delta_\phi +\ell -2)^2}{(d+2 \ell -4) (d+2 \ell -2)}\;,\\
{}&&C=\frac{(\Delta -1) (-d+\Delta +2) (\Delta +\ell )^2 (-d+\Delta +2 \Delta_\phi +\ell )^2}{4 (2 \Delta -d) (-d+2 \Delta +2) (\Delta +\ell -1) (\Delta +\ell +1)}\;,\label{ABCDE}\\
\nonumber {}&&D=\frac{(1-\Delta) \ell  (d-\Delta -2) (d+\ell -3) (d-\Delta +\ell -2)^2 (2 d-\Delta -2 \Delta_\phi +\ell -2)^2}{4 (d-2 \Delta -2) (d-2 \Delta ) (d+2 \ell -4) (d+2 \ell -2) (d-\Delta +\ell -3) (d-\Delta +\ell -1)}\;,\\
\nonumber {}&&E=\frac{1}{2} \left(-d \Delta +4 d \Delta_\phi +d \ell +\Delta ^2-4 \Delta_\phi ^2+\ell ^2-2 \ell \right)\;.
\end{eqnarray}
For simplicity, we will set $\ell=0$ for the l.h.s. of (\ref{crossedCasimir}). From the coefficients (\ref{ABCDE}) we find no unphysical negative spins arise, and only $\ell=0$ and $\ell=1$ appears on the r.h.s.. Using (\ref{Mellinscalar}), (\ref{Mellinvec1}) and crossing symmetry, we can write down the exchange Mellin amplitudes in the t-channel. Translating the Casimir operator according to (\ref{basicdifference}), it is straighforward although tedious to verify that the relation (\ref{crossedCasimirforW}) holds.

Let us now show an example of (\ref{crossedCasimir}) in position space where $\Delta_i=3$, $\Delta=4$. The t-channel scalar exchange reads 
\begin{equation}
\tilde{W}^{(d),t}_{4,0}=\frac{6U^2}{x_{12}^6x_{34}^6}\bar{D}_{2332}\;.
\end{equation}
On the r.h.s., we expect also vector exchange with dimensions $3$ and $5$
\begin{equation}
\begin{split}
\tilde{W}^{(d),t}_{3,1}={}&-\frac{3U^3}{2x_{12}^6x_{34}^6}\big((-\bar{D}_{1342}+\bar{D}_{1432}+\bar{D}_{2341}-U\bar{D}_{2431})\\
{}&-\frac{d-10}{d-8}(\bar{D}_{2343}-\bar{D}_{2433}-\bar{D}_{3342}+U\bar{D}_{3432})\big)\;,
\end{split}
\end{equation}
\begin{equation}
\tilde{W}^{(d),t}_{5,1}=\frac{15U^3}{4x_{12}^6x_{34}^6}(\bar{D}_{2343}-\bar{D}_{2433}-\bar{D}_{3342}+U\bar{D}_{3432})\;,
\end{equation}
which are related to (\ref{W31}) and (\ref{W51}) by crossing. To proceed, we decompose each $\bar{D}$-function above into the basis spanned by $1$, $\log U$, $\log V$ and $\Phi(U,V)$
\begin{equation}\label{inbasis}
F(U,V)=f_1+f_{\log U}\log U+f_{\log V}\log V+f_{\Phi} \Phi(U,V)\;,
\end{equation}
where $\Phi(U,V)$ is the scalar one-loop box function in four dimensions. This is possible because of the following differential recursion relations (see, {\it e.g.}, \cite{Arutyunov:2002fh})
\begin{equation}
\begin{split}
\bar{D}_{\Delta_1+1,\Delta_2+1,\Delta_3,\Delta_4}&=-\partial_U \bar{D}_{\Delta_1,\Delta_2,\Delta_3,\Delta_4}\;,\\
\bar{D}_{\Delta_1,\Delta_2,\Delta_3+1,\Delta_4+1}&=(\Delta_3+\Delta_4-\Sigma-U\partial_U )\bar{D}_{\Delta_1,\Delta_2,\Delta_3,\Delta_4}\;,\\
\bar{D}_{\Delta_1,\Delta_2+1,\Delta_3+1,\Delta_4}&=-\partial_V \bar{D}_{\Delta_1,\Delta_2,\Delta_3,\Delta_4}\;,\\
\bar{D}_{\Delta_1+1,\Delta_2,\Delta_3,\Delta_4+1}&=(\Delta_1+\Delta_4-\Sigma-V\partial_V )\bar{D}_{\Delta_1,\Delta_2,\Delta_3,\Delta_4}\;,\\
\bar{D}_{\Delta_1,\Delta_2+1,\Delta_3,\Delta_4+1}&=(\Delta_2+U\partial_U+V\partial_V )\bar{D}_{\Delta_1,\Delta_2,\Delta_3,\Delta_4}\;,\\
\bar{D}_{\Delta_1+1,\Delta_2,\Delta_3+1,\Delta_4}&=(\Sigma-\Delta_4+U\partial_U+V\partial_V )\bar{D}_{\Delta_1,\Delta_2,\Delta_3,\Delta_4}
\end{split}
\end{equation}
where $\Sigma=\frac{1}{2}(\Delta_1+\Delta_2+\Delta_3+\Delta_4)$, and also the identities satisfied by $\Phi(U,V)$
\begin{equation}\label{Phirecur}
\begin{split}
\partial_z\Phi&=-\frac{1}{z-\bar{z}}\Phi-\frac{1}{z(z-\bar{z})}\log V+\frac{1}{(-1+z)(z-\bar{z})}\log U\;,\\
\partial_{\bar{z}}\Phi&=\frac{1}{z-\bar{z}}\Phi+\frac{1}{\bar{z}(z-\bar{z})}\log V-\frac{1}{(-1+\bar{z})(z-\bar{z})}\log U\;.
\end{split}
\end{equation}
To simplify the action of the Casmir operator, it is more convenient to write it in terms of $z$ and $\bar{z}$
\begin{equation}
\mathrm{Cas}_s=-2(\mathbf{D}_z(a,b)+\mathbf{D}_{\bar{z}}(a,b))-2(d-2) \frac{z\bar{z}}{z-\bar{z}}\big((1-z)\frac{d}{dz}-(1-\bar{z})\frac{d}{d\bar{z}}\big)
\end{equation}
where 
\begin{equation}
\mathbf{D}_{z}(a,b)=(1-z)z^2\frac{d^2}{dz^2}-(1+a+b)z^2\frac{d}{dz}-abz\;.
\end{equation}
Using the relations (\ref{Phirecur}) recursively, we can again cast the action of the Casimir on the exchange diagram into the form of (\ref{inbasis}). We find that 
\begin{equation}
\mathrm{Cas}_s[\tilde{W}^{(d),t}_{4,0}]=2(2d-5)\tilde{W}^{(d),t}_{4,0}+4\tilde{W}^{(d),t}_{3,1}-\frac{4 (d-10) (d-6)}{5 (d-8)}\tilde{W}^{(d),t}_{5,1}+\tilde{W}_{extra}
\end{equation}
where 
\begin{equation}
\tilde{W}_{extra}=6(d-10)U^3\bar{D}_{3333}
\end{equation}
is a zero-derivative contact Witten diagram and can be absorbed into either $\tilde{W}^{(d),t}_{3,1}$ or $\tilde{W}^{(d),t}_{5,1}$. This confirms the prediction of (\ref{crossedCasimirforW}).

\vspace{0.3cm}
\noindent{\bf The irregular type}
\vspace{0.1cm}

\noindent Let us now study a few examples of recursion relations belonging to the irregular type. We start with a special case of (\ref{reF0}) with $\ell=0$ and $\Delta_i=\Delta_\phi$. The relation simplifies into
\begin{equation}
(U^{-1}-U^{-1}V)g^{(d)}_{\Delta,0}=-2 g^{(d)}_{\Delta-1,1}-\frac{\Delta ^2 (-d+\Delta +2)}{2 (\Delta +1) (2 \Delta -d) (-d+2 \Delta +2)}g^{(d)}_{\Delta+1,1}\;.
\end{equation}
Using the formulae (\ref{Mellinscalar}) and (\ref{Mellinvec1}) for the exchange Mellin amplitudes and (\ref{basicdifference}) for translating the cross ratio factor into an difference operator, we find that an extra term is needed in the Witten diagram relation
\begin{equation}\label{reF0sp}
U^{-1}(1-V)W^{(d)}_{\Delta,0}=-2 W^{(d)}_{\Delta-1,1}-\frac{\Delta ^2 (-d+\Delta +2)}{2 (\Delta +1) (2 \Delta -d) (-d+2 \Delta +2)}W^{(d)}_{\Delta+1,1}+F_{extra}\;.
\end{equation}
where 
\begin{equation}\label{F0Fextra}
F_{extra}=F_0+F_1+F_2\;,
\end{equation}
and
\begin{equation}\small
F_0=\frac{(V-1)}{x_{12}^{2\Delta_\phi}x_{34}^{2\Delta_\phi}}\frac{\Gamma (\Delta ) \Gamma \left(-\frac{d}{2}+\Delta +1\right) \Gamma \left(2 \Delta_\phi -\frac{d}{2}\right)U^{\Delta_\phi-1}\bar{D}_{\Delta_\phi\Delta_\phi\Delta_\phi\Delta_\phi}}{\Gamma \left(\frac{\Delta }{2}\right)^4 \Gamma \left(\frac{2\Delta_\phi -\Delta }{2}\right) \Gamma \left(\frac{2\Delta_\phi +2-\Delta }{2}\right) \Gamma \left(\frac{\Delta -d+2\Delta_\phi}{2} \right) \Gamma \left(\frac{-d+\Delta +2+2\Delta_\phi}{2} \right)}\;,
\end{equation}
\begin{equation}\small
F_1=\frac{\sqrt{\pi } 4^{1-\Delta_\phi } \Gamma (\Delta ) \Gamma (\Delta_\phi )^3 \Gamma \left(-\frac{d}{2}+\Delta +1\right) \Gamma \left(-\frac{d}{2}+2 \Delta_\phi +1\right)}{\Gamma \left(\frac{\Delta }{2}\right)^4 \Gamma \left(\Delta_\phi +\frac{1}{2}\right) \Gamma \left(-\frac{\Delta }{2}+\Delta_\phi +1\right)^2 \Gamma \left(\frac{1}{2} (-d+\Delta +2)+\Delta_\phi \right)^2}W^{(d)}_{2\Delta_\phi-1,1}\;,
\end{equation}
\begin{equation}\small
F_2=\frac{2^{\Delta -1} \Gamma \left(\frac{\Delta +1}{2}\right) \Gamma \left(-\frac{d}{2}+\Delta +1\right) \Gamma \left(2 \Delta_\phi -\frac{d}{2}\right)U^{\Delta_\phi}}{\sqrt{\pi } \Gamma \left(\frac{\Delta }{2}\right)^3 \Gamma \left(\frac{2\Delta_\phi-\Delta }{2}\right) \Gamma \left(\frac{2\Delta_\phi +2-\Delta }{2}\right) \Gamma \left(\frac{\Delta -d+2\Delta_\phi}{2} \right) \Gamma \left(\frac{-d+\Delta +2+2\Delta_\phi}{2} \right)}\frac{\bar{D}_{\Delta_\phi\Delta_\phi\Delta_\phi\Delta_\phi}}{x_{12}^{2\Delta_\phi}x_{34}^{2\Delta_\phi}}\;.
\end{equation}

Let us also test it in position space with a special example where $\Delta_\phi=3$, $\Delta=4$. The relevant diagrams has already been evaluated before, and 
\begin{equation}
\tilde{W}^{(d)}_{4,0}=\frac{6U^2}{x_{12}^6x_{34}^6}\bar{D}_{2233}\;.
\end{equation}
By decomposing into the basis (\ref{inbasis}), we find 
\begin{equation}
(U^{-1}-U^{-1}V)\tilde{W}^{(d)}_{4,0}=-2\tilde{W}^{(d)}_{3,1}+\frac{8 (d-9)}{5 (d-8)}\tilde{W}^{(d)}_{5,1}-6\frac{(1-V)}{x_{12}^6x_{34}^6}U^2\bar{D}_{3333}\;.
\end{equation}
This agrees with (\ref{reF0sp}) and (\ref{F0Fextra}). Note that $W^{(d)}_{2\Delta_\phi-1,1}$ in $F_1$ now coincide with $W^{(d)}_{\Delta+1,1}$, and $F_2$ appears if we change the choice of contact terms to go from $\tilde{W}^{(d)}_{\Delta,\ell}$ to $W^{(d)}_{\Delta,\ell}$.

Finally, let us discuss another interesting class of recursion relations which are the superconformal Ward identities. Four-point functions of one-half BPS operators in $d>2$, with R-symmetry $SO(n)$ and $n\geq 3$, obey the following superconformal Ward identities \cite{Dolan:2004mu} 
\begin{equation}\label{scfwardid}
(z\partial_z-\epsilon\alpha\partial_\alpha)\mathcal{G}(z,\bar{z};\alpha,\bar{\alpha})\big|_{\alpha=1/z}=0\;,\quad (\bar{z}\partial_{\bar{z}}-\epsilon\alpha\partial_{\alpha})\mathcal{G}(z,\bar{z};\alpha,\bar{\alpha})\big|_{\alpha=1/\bar{z}}=0
\end{equation}
where $\epsilon=\frac{d-2}{2}$, and $\alpha$, $\bar{\alpha}$ are R-symmetry cross ratios\footnote{When $n=3$ there is only one R-symmetry cross ratio.} analogous to $z$ and $\bar{z}$.\footnote{The two identities in (\ref{scfwardid}) are not independent because correlators are symmetric in $z$ and $\bar{z}$. Two more identities can be written down by replacing $\alpha$ with $\bar{\alpha}$, but they do not give new constraints because correlators are also symmetric in $\alpha$, $\bar{\alpha}$.} Superconformal blocks satisfy the above identities, and are given by a finite linear combination of bosonic conformal blocks 
\begin{equation}\label{scfCB}
\mathfrak{g}_{S}=\sum_{I\in S} R_I(\alpha,\bar{\alpha})g^{(d)}_{\Delta_I,\ell_I}(z,\bar{z})\;.
\end{equation}
Here $R_I(\alpha,\bar{\alpha})$ are polynomials eigenfunctions of the R-symmetry Casimir, and $I$ runs over components of the supermultiplet $S$ exchanged in the four-point function. Clearly, the  identities (\ref{scfwardid}) imply special relations for the non-supersymmetric conformal blocks in (\ref{scfCB}). To write (\ref{scfCB}) in terms of $U$ and $V$, we use the simple trick of taking the sum of the two identities  \cite{Zhou:2017zaw,Zhou:2018ofp}. As an example, we  focus on  3d $\mathcal{N}=8$ stress tensor supermultiplet exchange in the correlator of the same multiplet. The external operators are scalars with dimension 1, and the superconformal block is \cite{Chester:2018aca}
\begin{equation}
\mathfrak{g}^{3d,\mathcal{N}=8}_{stress}=(3-4\alpha-4\bar{\alpha}+8\alpha\bar{\alpha})g^{(3)}_{1,0}(z,\bar{z})+(\alpha+\bar{\alpha}-1)g^{(3)}_{2,1}(z,\bar{z})+\frac{3}{32}g^{(3)}_{3,2}(z,\bar{z})\;.
\end{equation}
Adding the two identities in (\ref{scfCB}) leads to the following equation
\begin{equation}\label{3dscfwardid}
\mathcal{D}_1g^{(3)}_{1,0}(z,\bar{z})+\mathcal{D}_2g^{(3)}_{2,1}(z,\bar{z})+\mathcal{D}_3 g^{(3)}_{3,2}(z,\bar{z})=0
\end{equation}
where 
\begin{equation}
\begin{split}
{}&\mathcal{D}_1=2U^{-1}(1+U-V)(1-2\bar{\alpha})+(1-U+7V+4\bar{\alpha}(U-1-3V))\partial_V\\
{}&\quad\quad\;\; +2(U-2+2V+4\bar{\alpha}(1-V))\partial_U\;,\\
{}&\mathcal{D}_2=(2U)^{-1}(V-U-1)+(-2V+\bar{\alpha}(U+V-1))\partial_V+(1-U-V+2U\bar{\alpha})\partial_U\;,\\
{}&\mathcal{D}_3=\frac{3}{32}(U+V-1)\partial_V+\frac{3}{16}U\partial_U\;.
\end{split}
\end{equation}
Translating $\mathcal{D}_i$ into difference operators $\hat{\mathcal{D}}_i$ in Mellin space introduces extra poles $s^{-2}t^{-1}$, while the corresponding exchange Mellin amplitudes\footnote{Here we rescaled the expressions from \cite{Zhou:2017zaw} and added some contact terms to simplify the expressions.} have only poles at $s=1+2m$
\begin{eqnarray}
&&\mathcal{M}^{(3)}_{1,0}=\sum_{m=0}^\infty\frac{2(-1)^{m+1}}{\pi ^{5/2} m! \Gamma \left(\frac{1}{2}-m\right)}\frac{1}{s-1-2m}\;,\\
&&\mathcal{M}^{(3)}_{2,1}=\sum_{m=0}\frac{8(-1)^{m+1}}{\pi ^{5/2} (2 m+1) \Gamma \left(\frac{1}{2}-m\right) \Gamma (m+1)}\frac{4-s-2t}{s-1-2m}\;,\\
\nonumber{}&&\mathcal{M}^{(3)}_{3,2}=\sum_{m=0}^\infty\frac{(-1)^{m+1}16 \Gamma \left(-m-\frac{3}{2}\right)}{3 \pi ^{5/2} m! \Gamma \left(\frac{1}{2}-m\right)^2} \frac{8 m (1-s)+4 (t-3) t+4 (u-3) u+4 m^2+19}{s-1-2m}\\
&&\quad\quad\quad+\frac{1}{\pi^2}(s-6)\;,
\end{eqnarray}
where $s+t+u=4$. Additional terms are thus needed to turn (\ref{3dscfwardid}) into a relation for Witten diagrams. It is not difficult to find 
\begin{equation}\label{AdS4Wid}
\hat{\mathcal{D}}_1\mathcal{M}^{(3)}_{1,0}+\hat{\mathcal{D}}_2\mathcal{M}^{(3)}_{2,1}+\hat{\mathcal{D}}_3\mathcal{M}^{(3)}_{3,0}=\mathcal{M}^{(3)}_{extra}
\end{equation}
where 
\begin{equation}
\mathcal{M}^{(3)}_{extra}=M_1+\bar{\alpha}\,M_2\;,
\end{equation}
\begin{equation}
M_1=\frac{2}{\pi^2}\left(\widehat{U^{-1}}[s-1]+\widehat{U^{-1}V}[1]-u\right)\;,\quad \quad M_2=-\frac{4}{\pi^2}\;.
\end{equation}
We can also rewrite (\ref{AdS4Wid}) in position space. The extra term $\mathcal{M}^{(3)}_{extra}$ corresponds to a linear combination of contact Witten diagrams with coefficients depending on the cross ratios. 

The non vanishing of $\mathcal{M}^{(3)}_{extra}$ is the statement that combinations of exchange Witten diagrams in one channel alone cannot solve the superconformal Ward identities. 
Moreover, one can also prove that adding contact terms with no more than two derivatives to the exchange amplitudes will not eliminate $\mathcal{M}^{(3)}_{extra}$. These facts are essential for the bootstrap methods  for holographic correlators \cite{Rastelli:2016nze,Rastelli:2017udc,Zhou:2017zaw,Rastelli:2017ymc,Zhou:2018ofp} to succeed, which claim that an ansatz of exchange and contact Witten diagrams in {\it all} channels can be uniquely solved by imposing the superconformal Ward identities (\ref{scfwardid}).

\bibliography{dimred} 
\bibliographystyle{utphys}

\end{document}